\documentstyle[12pt, graphicx]{article} 
\def\baselinestretch{1.3}
\def\no{\noindent}

\def\st{\scriptstyle}

\def\s{\sigma}
\def\st{\s^3}
\def\w{\omega}
\def\r{\rho}

\def\bea{\begin{eqnarray}}
\def\eea{\end{eqnarray}}

\def\gs{g_\s}
\def\gw{g_\w}
\def\gr{g_\r}

\def\gs2{g_\s^2}
\def\gw2{g_\w^2}
\def\gr2{g_\r^2}

\def\grp{g_\r^2/4\pi}

\def\ms{m_\s}

\def\mp{m_\pi}
\def\ms*{m_\s^*}

\def\msd{m_\s^2}

\def\mp2{m_\pi^2}
\def\mr2{m_\r^2}
\def\ms*2{m_\s^{*2}}

\def\bb{\bar b}
\def\cb{\bar c}
\def\pb{\bar \psi}
\def\p{\psi}

\def\O{^{16}O}
\def\Ca{^{40}Ca}
\def\Ca48{^{48}Ca}

\def \sa {\mbox{${^{16}O}$}}
\def \sb {\mbox{${^{40}Ca}$}}
\def \sc {\mbox{${^{48}Ca}$}}
\def \sd {\mbox{${^{90}Zr}$}}
\def \se {\mbox{${^{208}Pb}$}}

\def \mc  {\multicolumn{1}}
\def \mcd {\multicolumn{2}}
\def \mct {\multicolumn{3}}
\def \mcq {\multicolumn{4}}

\def \mc  {\multicolumn{1}}
\def \mcd  {\multicolumn{2}}

\newcommand{\rs}{\rho_s}
\newcommand{\sq}{\rho_{sq}}
\newcommand{\vq}{\rho_{vq}}
\newcommand{\Tq}{\rho_{Tq}}
\newcommand{\sn}{\rho_{sn}}
\newcommand{\vn}{\rho_{vn}}
\newcommand{\Tn}{\rho_{Tn}}
\renewcommand{\sp}{\rho_{sp}}
\newcommand{\vp}{\rho_{vp}}
\newcommand{\Tp}{\rho_{Tp}}

\newcommand{\Si}{\Sigma}
\newcommand{\dl}{\delta}
\newcommand{\ka}{\kappa}

\newcommand{\ga}{\gamma}

\newcommand{\la}{\lambda}
\newcommand{\ba}{\begin{eqnarray}&&}

\newcommand{\ea}[1]{\label{#1}\end{eqnarray}}
\newcommand{\be}{\begin{equation}}
\newcommand{\ee}[1]{\label{#1}\end{equation}}
\newcommand{\mf}[2]{\frac{\displaystyle#1}{\displaystyle#2}}
\newcommand{\pt}[3]{\bar{\psi}_i(1)\psi_{#1}(1)
\bar{\psi}_j(2)\psi_{#2}(2)
\bar{\psi}_k(3)\psi_{#3}(3)}
\newcommand{\qt}[3]{\bar{\psi}_i(1)\bar{\psi}_j(2)
\bar{\psi}_k(3)\psi_{#1}(3)
\psi_{#2}(2)\psi_{#3}(1)}
\newcommand{\pf}[4]{\bar{\psi}_i(1)\psi_{#1}(1)
\bar{\psi}_j(2)\psi_{#2}(2)
\bar{\psi}_k(3)\psi_{#3}(3)
\bar{\psi}_m(4)\psi_{#4}(4)}
\newcommand{\qf}[4]{\bar{\psi}_i(1)\bar{\psi}_j(2)
\bar{\psi}_k(3)\bar{\psi}_m(4)\psi_{#1}(4)
\psi_{#2}(3)\psi_{#3}(2)\psi_{#4}(1)}

\title{\bf Description of nuclear systems within the relativistic 
Hartree-Fock method with zero range self-interactions of the 
scalar field}
   
\author{S. Marcos$^1$, L. N. Savushkin$^2$, V. N. Fomenko$^3$, 
M. L\'opez-Quelle$^4$\\ and R. Niembro$^1$ \\ 
$^1$ \small \em Departamento de F\'\i sica Moderna,
\small \em  Universidad de Cantabria,  
E-39005 Santander, Spain       \\ 
$^2$ \small \em Department of Physics,
\small \em St. Petersburg University for Telecommunications,\\ 
\small \em 191065 St. Petersburg, Russia\\
$^3$ \small \em Department of Mathematics,
\small \em St. Petersburg University for Railway Engineering,\\ 
\small \em 190031 St. Petersburg, Russia\\
$^4$ \small \em Departamento de F\'\i sica Aplicada, 
\small \em Universidad de Cantabria,  E-39005 Santander, Spain}

\date{ }

\begin{document} 
\maketitle 
\begin{abstract} 
An exact method is suggested to treat the 
nonlinear self-interactions (NLSI) 
in the relativistic Hartree-Fock (RHF) approach for nuclear systems.
We consider here the NLSI constructed from 
the relativistic scalar nucleon 
densities and including products of six and eight fermion fields. This 
type of NLSI corresponds to the zero range limit of the standard cubic and 
quartic self-interactions of the scalar field. 
The method to treat the NLSI uses the Fierz transformation, 
which enables one to express the exchange (Fock) components in terms of the direct (Hartree) ones. 
The method is applied to nuclear matter and finite nuclei. 
It is shown that, in the RHF formalism, the NLSI,
which are explicitly isovector-independent, generate scalar,
vector and tensor nucleon self-energies strongly density-dependent. 
This strong isovector structure of the self-energies is
due to the exchange terms of the RHF method.
Calculations are carried out with a
parametrization  containing  five  free  parameters. The model allows
a description of both types of systems 
compatible with experimental data.
\end{abstract}  

PACS number(s): 21.30.Fe, 21.10.-k, 21.60.Jz, 21.65.+f.


Short title: Relativistic Hartree-Fock description of nuclear systems.
%\newpage

\bigskip
%\section{INTRODUCTION}
\section{Introduction}

The Relativistic Hartree-Fock (RHF) approach for finite nuclei has been developed 
in Refs. [1-8] (see also references therein) for the so-called 
linear models, which are characterized by linear field equations. 
In that approach, the contribution of the pion degrees of freedom 
(single pion exchange) is taken into account explicitly. 
In Ref. [2] it is shown, in particular, that the pseudovector coupling of pions to nucleons 
is more preferable than the pseudoscalar one in the nuclear structure context.
In Ref. [9], certain important features concerning the spin-orbit interation are 
directly related to the pion contribution. 
Thus, the incorporation of pions into the model 
represents one of the main advantages of the RHF method in comparison 
to the relativistic Hartree approach. 
However, in the papers mentioned above the  nonlinear 
self-interactions (NLSI) of the mesonic fields have 
not been taken into account. Including nonlinear self-interaction
terms corresponds to one of the possibilities to account for the three- and 
four-body forces in the nuclear structure calculations. Different types 
of self-interaction Lagrangians have been considered in literature up to 
now. Initially, the scalar field self-interactions have been introduced 
involving $\st$- and $\s^4$-terms, where $\s$ corresponds to the nuclear 
scalar field [11]. This type of self-interactions has been shown to play a 
very essential role in the relativistic Hartree calculations [12,13] to get, 
for example, the proper value of the compressibility modulus ($K$).
One of the main problems of the RHF approach is that it brings about  a high value of $K$.
Thus, the inclusion of the NLSI terms in the RHF famework could solve this problem too.
To work out the exact RHF equations with the NLSI
is a complicated task not solved up to now. 
In Ref. [14], an approximate method to take into account self-interactions 
of the $\st$- and $\s^4$-type
in the RHF procedure has been suggested (see Ref. [15] for other type of NLSI). 
This method involves a simple idea based on the 
inclusion of the nonlinear terms appearing in the equation for the $\sigma$ 
field, together with the scalar meson mass $m_\sigma$, into a scalar 
meson effective mass $m^*_\sigma$, which replaces $m_\sigma$ in the 
corresponding meson propagator in the nuclear medium.
Let us mention also that in Ref. [10] the authors present a procedure which 
allows one to evaluate the contribution of the Fock terms in a truncation scheme, 
including self-interactions of the scalar meson field, and illustrate their method in the 
case of infinite nuclear matter.

In the present paper, we study the properties of a Lagrangian including, 
beside the exchange of $\s$, $\w$, $\pi$ and $\rho$ mesons between nucleons,
the NLSI of the scalar field in the zero range limit (ZRL),
in the framework of the RHF approximation.
The ZRL allows one to express the exchange (Fock) terms for the NLSI 
explicitly via the direct (Hartree) terms in an exact way.

The paper is arranged as follows. In Sect. II, the general formalism is 
presented extensively: the Hartree-Fock contributions of the
NLSI are calculated (in the ZRL without any approximation), 
both to the total energy of the system and to the nucleon self-energies, 
the expressions obtained being valid for nuclear matter (NM) and finite nuclei. 
For both cases, the results  are shown in Sect. III. Finally, 
the conclusions are drawn in Sect. IV.

%\section{GENERAL FORMALISM}
\section{General formalism}

\centerline{\it A. The effective Lagrangian}

The effective Lagrangian of the model considered in the present paper 
involves interactions of the nucleons via the exchange of mesons with the space-time 
transformation properties of the scalar $\s$, vector (both isoscalar 
$\w$, and isovector $\r$), and pseudoscalar $\pi$ fields. 
It contains a "linear" part, which generates linear field equations 
identical to the corresponding linear part of Ref. [14], 
and it will not be reproduced here. Let us mention only that 
the pion field is coupled to nucleons through a pseudovector coupling [2].
The current Lagrangian involves, however, a
self-interaction part that is somewhat different from 
its homologous part of Ref. [14]. It takes the form:

\bea
U_{SI}=-\frac1{3}b{(\frac{g_\s}{\msd})}^3(\pb\p)^3
+\frac1{4}c(\frac{g_\s}{\msd})^4(\pb\p)^4,
\label{en1}
\eea

\no
where $\p$ is the fermion field 
operator, $b$ and $c$ are the coupling constants of the nonlinear terms,
$g_\s$ and $m_\s$ are, respectively, the scalar meson coupling constant and mass. 
We shall utilize also the dimensionless coupling constants $\bb=\frac{b}{g^3_\s M}$ and 
$\cb=\frac{c}{{g^4_\s}}$, $M$ being the nucleon mass.

 Notice that the NLSI given in Eq. (1) coincide with the conventional scalar 
field NLSI used in Ref. [14] in the ZRL of the scalar field, i.e., 
when the term with $m_{\sigma}^2$ in the equation of motion of the scalar field dominates
over the Laplacian and nonlinear terms. This is what happens for nuclear systems at 
small densities and with a smooth surface. Although, strictly speaking, 
real finite nuclei do not satisfy these conditions, for values of 
$m_{\sigma} \stackrel{>}{\sim} 500$ MeV, we can consider that this approximation 
in effective Lagrangians is acceptable.
The parameters $b$, $c$, $\bb$ and $\cb$ have the same units as in Ref. [14].

Let us mention also that the NLSI given by Eq. (1) have just the same 
structure as the respective components of the Lagrangian appearing in 
the point coupling model and containing higher order terms in the 
fermion fields [17,18]. The inclusion of this kind of terms in the Lagrangian 
can be also justified by the necesity to introduce an extra density 
dependence in the Dirac-Brueckner-Hartree-Fock calculations to allow a simultaneous 
fit to the NN phase shifts and the nuclear matter equilibrium point [19].

\bigskip
\centerline{\it B. The Hartree-Fock equations}

To get the RHF equations corresponding to our effective Lagrangian,
we closely follow the Refs. [3,14], restricting ourselves to the static 
approximation for the meson fields. The nucleon field $\p$ is 
expanded into a complete set of stationary single-particle spinors 
$\{\p_\alpha (x)e^{-iE_\alpha t}\}$ and we consider the tree approximation.

The Dirac equation including exchange terms can be obtained by minimizing
the total energy of the system, which 
is given by the expectation value of the total Hamiltonian 
in the space of Slater determinants $\Pi_\alpha a_\alpha^+|0>$, where $a_\alpha^+$ is a creation 
operator for a nucleon in the state $\alpha$.

\bigskip
\centerline{\it B.1  Contributions of the NLSI terms to the total energy of the system}

The contribution to the energy of the linear part of the Lagrangian, of finite range,
is taken directly from Ref. [3]. It includes, of course, the exchange terms.
Here, we shall concentrate in the technique of calculating 
direct (Hartree) and exchange (Fock) components of the contribution 
to the energy of the self-interaction term $U_{SI}$ given by Eq. (1).

In doing this, we obtain for the mean value of the cubic component of $U_{SI}$

\bigskip
\begin{eqnarray}
<0|(\pb\p)^3|0>&&=\langle0|\sum_{ijk}\sum_{i'j'k'}\qt{k'}{j'}{i'}a_i^+a_j^+a_k^+a_{k'}a_{j'}a_{i'}
|0\rangle\nonumber\\
&&=\sum_{ijk}\pt ijk\nonumber\\
&&-3\sum_{ijk}\pt ikj\\
\label{en2}
&&+2\sum_{ijk}\pt kij,\nonumber
\end{eqnarray}

\no
and for the mean value of the quartic component 
\begin{eqnarray}
<0|(\pb\p)^4|0>&&=\langle0|\sum_{ijkm}\sum_{i'j'k'm'}\qf{m'}{k'}{j'}{i'}\nonumber\\
&&\times a_i^+a_j^+a_k^+a_m^+a_{m'}a_{k'}a_{j'}a_{i'}|0\rangle\nonumber\\
&&=\sum_{ijkm}\pf ijkm\nonumber\\
&&-6\sum_{ikjm}\pf ikjm\nonumber\\
&&+8\sum_{ikjm}\pf imjk\\
\label{en3}
&&-6\sum_{ikjm}\pf mijk\nonumber\\
&&+3\sum_{ikjm}\pf jimk.\nonumber
\end{eqnarray}

Let us mention that the space coordinates in Eqs.(2) and (3) coincide for all spinors in the ZRL 
and that the subscripts $i,j,k$ in the last three lines of Eq. (2) and $i,j,k,m$ 
in the last four lines of Eq. (3) run over single-particle occupied states. 
Let us emphasize also that the sum of coefficients of the exchange (Fock) 
terms and of the direct (Hartree) term in Eqs. (2) and (3) is equal to zero, as it should be.

We define the scalar ($\rho_{sq}$), vector ($\rho_{vq}$) and tensor ($\rho_{Tq}$) 
densities for neutrons ($q=n$) or protons ($q=p$) in the usual way:

\begin{eqnarray}
\rho_{sq}(1)=&&\sum_{i(q\;states)}\bar{\psi_i}(1)\psi_i(1),\nonumber\\
\rho_{vq}(1)=&&\sum_{i(q\;states)}\bar{\psi_i}(1)\gamma^0\psi_i(1),\\
\label{den}
\rho_{Tq}(1)=&&\sum_{k=1}^3\sum_{i(q\;states)}\bar{\psi_i}(1)\sigma^{0k}\psi_i(1)
n^k=i\sum_{i(q\;states)}\bar{\psi_i}(1)\gamma^0\vec{\gamma}\cdot\vec n\psi_i(1),
\nonumber
\end{eqnarray}

\no
where the subscript $i$ runs over all the occupied states of a nucleon of type $q$. Here, 
$\vec n$ is the unit vector along the radial direction.

The scalar, vector and tensor total densities are $\rho_s=\rho_{sn}+\rho_{sp}$,
$\rho_v=\rho_{vn}+\rho_{vp}$ and $\rho_T=\rho_{Tn}+\rho_{Tp}$, respectively.

Using the Fierz transformation [16], we can write for a fermion system:

\begin{eqnarray}
\sum_{a,b}(\bar{\psi_a}\psi_b)(\bar{\psi_b}\psi_a)&&=
\frac{1}{4}\sum_{a,b}[(\bar{\psi_a}\psi_a)(\bar{\psi_b}\psi_b)
+(\bar{\psi_a}\gamma_5\psi_a)(\bar{\psi_b}\gamma_5\psi_b)
+(\bar{\psi_a}\gamma_\mu\psi_a)(\bar{\psi_b}\gamma_\mu\psi_b)\nonumber\\
&&-(\bar{\psi_a}\gamma_5\gamma_\mu\psi_a)(\bar{\psi_b}\gamma_5\gamma_\mu\psi_b)+
\frac{1}{2}(\bar{\psi_a}\sigma_{\mu\nu}\psi_a)(\bar{\psi_b}\sigma^{\mu\nu}\psi_b)].
\end{eqnarray}

The quantities $\sum_{a,b,c}(\bar{\psi_a}\psi_b)(\bar{\psi_b}\psi_c)(\bar{\psi_c}\psi_a)$ and
$\sum_{a,b,c,d}(\bar{\psi_a}\psi_b)(\bar{\psi_b}\psi_c)(\bar{\psi_c}\psi_d)(\bar{\psi_d}\psi_a)$
can also be written in a similar way, although they involve much more terms and 
will not be given here.

In order to write down the quantities $<0|(\pb\p)^3|0>$ and $<0|(\pb\p)^4|0>$ in terms of the
nucleon densities defined in Eqs. (4), we take into account the following relation:

\begin{equation}
\sum_i\bar{\psi}_i\sigma^{\mu\nu}\psi_i=\sum_{k=1}^3
(\delta_{\mu0}\delta_{\nu k}-\delta_{\mu k}\delta_{\nu0})\rho_Tn^k.
\label{rel1}
\end{equation}

From this relation, one gets two more useful equations

\begin{equation}
\sum_i\bar{\psi}_i\sigma^{\mu\nu}\psi_i\sum_i\bar{\psi}_i\sigma_{\mu\nu}\psi_i=
-2\rho_T^2
\label{rel2}
\end{equation}
and
\begin{equation}
\sum_i\bar{\psi}_i\sigma^{\mu\nu}\psi_i\sum_i\bar{\psi}_i\sigma_{\mu\ka}\psi_i=
\rho_T^2\left[(1-\delta_{\nu0})(1-\delta_{\ka0})n^{\nu}n_{\ka}
-\delta_{\nu0}\delta_{\ka0}\right].
\label{rel3}
\end{equation}

Finally, from this last equation one obtains
\begin{equation}
\sum_i\bar{\psi}_i\sigma^{\mu\nu}\psi_i\sum_i\bar{\psi}_i\sigma_{\mu\ka}\psi_i
\sum_i\bar{\psi}_i\sigma_{\la\nu}\psi_i\sum_i\bar{\psi}_i\sigma^{\la\ka}\psi_i=
2\rho_T^4.
\label{rel4}
\end{equation}

Having in mind that the space coordinates in Eqs. (2) and (3) coincide 
for all spinors in the ZRL, we get from Eq. (2) for the cubic term

%\newpage

\begin{eqnarray}
\langle0|(\pb\p)^3|0\rangle
&&=\rho^3_s-\frac34\rho_s\left[\rho_{sn}^2+\rho_{sp}^2+
\rho_{vn}^2+\rho_{vp}^2-(\rho_{Tn}^2+\rho_{Tp}^2)\right]\nonumber
%\label{end3p}
\\
&&+\frac18\left[\rho_{sn}^3+\rho_{sp}^3+
3(\rho_{sn}\rho_{vn}^2+\rho_{sp}\rho_{vp}^2)-
3(\rho_{sn}\rho_{Tn}^2+\rho_{sp}\rho_{Tp}^2)\right],
\end{eqnarray}

\no
and from Eq. (3) for the quartic term

\begin{eqnarray}
\langle0|(\pb\p)^4|0\rangle
&&=\rho_s^4-\frac32\rho_s^2\left[\rho_{sn}^2+\rho_{sp}^2+
\rho_{vn}^2+\rho_{vp}^2-(\rho_{Tn}^2+\rho_{Tp}^2)\right]
\nonumber
%\label{end4p}
\\
&&+\frac12\rho_s\left[\rho_{sn}^3+\rho_{sp}^3+
3(\rho_{sn}\rho_{vn}^2+\rho_{sp}\rho_{vp}^2)-
3(\rho_{sn}\rho_{Tn}^2+\rho_{sp}\rho_{Tp}^2)\right]\nonumber\\
&&-\frac3{32}\left[\rho_{sn}^4+\rho_{sp}^4
+6(\rho_{sn}^2\rho_{vn}^2+\rho_{sp}^2\rho_{vp}^2)
-6(\rho_{sn}^2\rho_{Tn}^2+\rho_{sp}^2\rho_{Tp}^2)\right.\\
&&\left.+\rho_{vn}^4+\rho_{vp}^4
-2(\rho_{vn}^2\rho_{Tn}^2+\rho_{vp}^2\rho_{Tp}^2)
+\rho_{Tn}^4+\rho_{Tp}^4\right]\nonumber\\
&&+\frac3{16}\left(\rho_{sn}^2+\rho_{sp}^2+\rho_{vn}^2+\rho_{vp}^2
-\rho_{Tn}^2-\rho_{Tp}^2\right)^2.\nonumber
\end{eqnarray}

Let us mention that the contributions given by Eqs. (10) and (11) include 
the densities $\rho_s$, $\rho_v$ and $\rho_T$ but they do not include the contributions of 
pseudoscalar $\rho_P$- and axial vector $\rho_A$-densities for parity reasons.

From Eqs. (10) and (11), it can be easily seen that the exchange (Fock) 
contributions to the energy of the system 
corresponding to both the cubic and quartic terms are very essential.
Actually, only the $\rs^3$-term in Eq. (10) 
and the $\rs^4$-term in Eq. (11) originate from the direct (Hartree) 
contributions, all the other terms in Eqs. (10) and (11) arise from the 
Fock contributions, and they have a strong isovector structure 
(ISVS)\footnote{We consider that a quantity has ISVS if it depends on the difference 
$\rho_n-\rho_p$ (or $\rho_{sn}-\rho_{sp}$).}.

\bigskip
\centerline{\it B.2 Contributions of the  NLSI terms to the nucleon self-energy}

The contribution of the NLSI terms of the third (3) and forth (4) order
to the self-energy of a nucleon of type $q$ [$\hat \Si_q^{(3,4)}$] 
can be extracted from the following equation:

\be
\hat \Si^{(3,4)}_q\psi_q=\frac{\dl}{\dl\bar{\psi_q}}U_{SI}^{(3,4)}=
[\Si^{(3,4)}_{sq}+\Si^ {(3,4)}_{vq}\ga^0+i\Si^{(3,4)}_{Tq}\ga^0\vec{\ga}\cdot\vec{n}]\psi_q.
\ee{self}
From Eq. (12), it is seen that the NLSI terms $U_{SI}$ 
give a contribution to the total nucleon self-energy 
of the same structure as the linear components of the Lagrangian. 
Thus, each component ($\Si_{iq}$) of the total self-energy can be written as
\be
\Si_{iq}=\Si_{iq}^{linear}+\Si_{iq}^{(3,4)},
\ee{}

\no
where $i=s,0,T$ specifies the scalar, time component of the vector, and 
tensor self-energies. One can look in Refs. [3,14]
for further details related to $\Si_{iq}^{linear}$. 

Taking into account that
\be
\frac{\dl}{\dl\bar{\psi}_i}\rho_s=\psi_i,\quad
\frac{\dl}{\dl\bar{\psi}_i}\rho_v=\ga^0\psi_i, \hbox{and}\quad 
\frac{\dl}{\dl\bar{\psi}_i}\rho_T=i\ga^0\vec{\ga}\cdot\vec n\psi_i,
\ee{der}

\no
one obtains for the contribution of the cubic terms of $U_{SI}$ to the self-energy components

\begin{eqnarray}
\Si_{sq}^{(3)}=3\rs^2-\frac34(\sn^2+\sp^2+\vn^2+\vp^2-\Tn^2-\Tp^2)
-\frac38(4\rs\sq-\sq^2-\vq^2+\Tq^2),\ea{Z3s}

\begin{eqnarray}
\Si_{0q}^{(3)}=\mf34\vq(\sq-2\rs),\qquad
\ea{Z30}

\begin{eqnarray}
\Si_{Tq}^{(3)}=\mf34\Tq(2\rs-\sq).\qquad
\ea{Z3T}

For the contribution of the quartic terms of $U_{SI}$ to the self-energy components, one has

\begin{eqnarray} 
\Si_{sq}^{(4)}&&=4\rs^3+\frac34(\sq-4\rs)\left(\sn^2+\sp^2+\vn^2+\vp^2-\Tn^2-\Tp^2\right)\nonumber\\
&&+\frac12(\sn^3+\sp^3)+\frac32(\sn\vn^2+\sp\vp^2)
-\frac32(\sn\Tn^2+\sp\Tp^2)\label{Z4s}\\
&&-3\rs^2\sq+\frac32\rs(\sq^2+\vq^2-\Tq^2)
-\frac3{8}\sq(\sq^2+3\vq^2-3\Tq^2),\nonumber
\end{eqnarray}

\begin{eqnarray}
\Si_{0q}^{(4)}=3\vq\left[\rs\sq-\rs^2
-\frac1{8}(3\sq^2+\vq^2-\Tq^2)+\frac14(\sn^2+\sp^2+\vn^2+\vp^2-\Tn^2-\Tp^2)\right],
\label{Z40}
\end{eqnarray}

\begin{eqnarray}
\Si_{Tq}^{(4)}=3\Tq\left[\rs^2-\rs\sq+\frac1{8}(3\sq^2+\vq^2-\Tq^2)
-\frac14(\sn^2+\sp^2+\vn^2+\vp^2-\Tn^2-\Tp^2)\right].
\label{Z4T}
\end{eqnarray}

From the previous publications [3,14], it is clear that the isovector 
structure of the Hartree-Fock solutions is strongly determined by the 
contribution of the $\pi$- and $\rho$-mesons.
The present results, show that the isovector-independent NLSI
also make essential contributions to the ISVS of the energy and self-energies of the system.

From Eqs. (12,13), it follows that the quantities 
$\Si_{sq}^{(3,4)}$, $\Si_{0q}^{(3,4)}$ and $\Si_{Tq}^{(3,4)}$ 
enter the RHF Dirac equation in the same manner as the self-energies 
$\Si_{sq}$, $\Si_{0q}$ and $\Si_{Tq}$, produced by linear interactions, do. 
However, the $\Si^{(3,4)}_{iq}$-components involve a strong density dependence, 
and, what is more surprising, a strong ISVS.
The densities $\rho_{sq}$, $\rho_{vq}$ and 
$\rho_{Tq}$ can be calculated in the following way

\begin{eqnarray}
\rho_{sq}(r)=&&\frac1{4\pi r^2}\sum_{i(q\;states)}\left[G_{i}^2(r)-F_{i}^2(r)\right],\nonumber\\
\rho_{vq}(r)=&&\frac1{4\pi r^2}\sum_{i(q\;states)}\left[G_{i}^2(r)+F_{i}^2(r)\right],\\
\label{dens}
\rho_{Tq}(r)=&&\frac1{2\pi r^2}\sum_{i(q\;states)}G_{i}(r)\cdot F_{i}(r),\nonumber
\end{eqnarray}

\no
where $G_{i}(r)/r$ and $F_{i}(r)/r$ are the radial functions of the upper and 
lower components of the nucleon Dirac spinor.
Let us mention that Eqs. (10, 11, 15-17) are valid for NM and finite nuclei. 

\newpage
%\section{NUMERICAL RESULTS}
\section{Numerical results}

\centerline{\it A. Nuclear Matter}

We shall start the discussion of our results for symmetric NM, where $\rho_T=0$, 
$\rho_{s,p}=\rho_{s,n}$ and $\rho_{v,p}=\rho_{v,n}$.
As is mentioned above, in the present paper we follow closely the Lagrangian 
of Ref. [14] corresponding to the set HFSI  and, 
in our calculations, we have fixed the same parameters as in this reference. Thus, the bare 
nucleon mass and the $\pi$, $\w$ and $\r$ meson masses have been taken 
to be equal to their empirical values: $M=939$ MeV, $m_\pi=138$ MeV, 
$m_\w=783$ MeV, and $m_\r=770$ MeV. As for the $f_\pi$ and $g_\r$ coupling 
constants, we have chosen the experimental values $f^2_\pi/4\pi=0.08$
and $\grp=0.55$. The ratio $f_\r/g_\r=3.7$ is taken in accordance with 
the vector dominance model. Then, we are left with the following five 
free parameters: $m_\s$, $g_\s$, $g_\w$, $b$ and $c$, which are to be 
adjusted to reproduce some observables for NM and finite nuclei 
in a similar manner as in Ref. [14].

In Fig. 1, we present the result of the RHF calculation 
of the energy per particle ($E/A$) 
for NM as a function of $\r_v$ for three cases: 
1) The curve HFSI corresponds to the results 
obtained by the method suggested in Ref. [14] to take into account NLSI.
2) The curve ZRL$^*$[b,c(HFSI)]  corresponds to the results 
obtained with the Lagrangian and the method proposed in this paper, 
taking for the parameters $m_{\sigma}$, $b$ and $c$ 
the same values as in the HFSI approximation [14], $g_\s$ and $g_\w$ being chosen 
to get the saturation point at the same density, 
$\r_0=0.14$ fm$^{-3}$, as in the HFSI set. 
We obtain in this case the compressibility modulus $K=319$ MeV.
3) The curve ZRL[$g_\s$, $g_\w$, b,c(HFSI)] corresponds
to the results obtained for the same set of 
parameters as in the HFSI set of Ref. [14], the NLSI being treated in the 
ZRL by the method suggested in the present paper (the 
saturation point is achieved at $\sim 0.113$ fm$^{-3}$ in this 
case)\footnote{In the limit of small densities, where the contribution of the 
self-interaction terms becomes negligible, the HFSI- and ZRL-curves coincide.}.
To check the reliability of the ZRL approximation, 
we have carried out calculations of the $E/A$ vs $\r_v$-curves in 
the Hartree approach also. The results of the calculations are presented 
in Fig. 2. The ZRL approximation introduces appreciable differences 
with respect to the exact calculation for densities around the saturation value.

Considering $g_\s, g_\w$, $b$ and $c$ as fitting parameters,
it is easy to determine a set with a compressibility modulus $K$ 
equal to a given reasonable phenomenological value.
In Fig. 3, the energy per particle E/A as a function of $\r_v$ is represented 
for the model considered in the present paper (ZRL) for three sets of parameters. 
In all sets the scalar mass and one coupling parameter are chosen with the same values 
as in the HFSI set of Ref. [14], 
whereas the other three coupling parameters are chosen to get the same 
values of $\rho_0$ and $E(\rho_0)/A$ in NM 
as in the HFSI set (i.e. $\r_0$=0.14 fm$^{-3}$, 
$E/A=-15.75$ MeV) and $K$=250 or 275 MeV:

1) ZRL$^{c1}$ set: $\cb$ is fixed to the value 
($\cb=-0.01461$) obtained in the HFSI set, 
while $\bb=-0.011533$, $\frac{\gs2}{4\pi}=3.0621$, $\frac{\gw2}{4\pi}=7.849$ 
generate the required saturation conditions of NM with $K=275$ MeV.

2) ZRL$^{b1}$ set: $\bb$ is fixed to the value 
($\bb=-0.006718$) obtained in the HFSI set, 
while $\cb=0.012475$, $\frac{\gs2}{4\pi}=3.3039$, $\frac{\gw2}{4\pi}=8.9346$ 
generate the required saturation point of NM with $K=275$ MeV.

3) ZRL$^{b2}$ set: $\bb$ is fixed to the value 
($\bb=-0.006718$) obtained in the HFSI set, 
while $\cb=0.042561$, $\frac{\gs2}{4\pi}=3.1715$, $\frac{\gw2}{4\pi}=8.4777$ 
generate the required saturation point of NM with $K=250$ MeV.

One could also try to get $K=250$ MeV keeping the value of $c$ as in the 
HFSI set, however, we could not find solutions in this case. From Fig. 3, 
one can see that appreciable differences between the three models
appear only at densities larger than the saturation one. Thus, a similar 
description of finite nuclei can be expected for these three sets.

Our next step is to find an adequate parameterization for NM and finite nuclei. 
To do that, we can follow the procedure used in Ref. [14].
In the present paper, the values of $g_\s$, $b$, $c$ are determined by reproducing the saturation 
conditions of symmetric NM 
for $\r_0$, $E/A$ and a reasonable value of the compressibility modulus.
In this work we have fitted two sets, ZRL1 and ZRL2, corresponding to the compressibility modulus 
250 MeV and 275 MeV, respectively. 
Calculations for finite nuclei put some extra constraints on the 
values of $m_\s$ and $g_\w$ ($m_\s$ is adjusted to get the 
experimental $r.m.s.$ charge radius of the $\O$ and 
the value of $g_\w$ is chosen to get reasonable values of spin-orbit splittings). 
Alternatively, one can use finite nuclei data to fit directly the 
free parameters of the model.

The values of the parameters chosen in the present paper for the ZRL1 and ZRL2 sets
and for the HFSI set of Ref. [14], 
as well as some calculated NM properties, are given in Table I.
One can appreciate that the scalar meson mass $m_\s$ needed to get a 
good description of the surface nuclear properties is considerably 
larger in the ZRL1 and ZRL2 sets than in the HFSI one. 
The parameters $\bb$, $\cb$ are quite different in the sets ZRL1 and ZRL2 
as compared with those of the set HFSI.
This is due to the strong Fock terms contribution of the NLSI in sets ZRL1 and ZRL2.

Fig. 4 shows for the ZRL1 set the values of $E/A$ as a function of $\r_v$ for  
different values of the asymmetry parameter $\delta$ defined by the 
following equation:

\begin{eqnarray}
\delta=\frac{\r_{v,n}-\r_{v,p}}{\r_{v,n}+\r_{v,p}}\label{del}
\end{eqnarray}

It can be appreciated in this figure a strong shift of the 
equilibrium density $\r_0(\delta)$ towards smaller values as $\delta$ increases 
and also that the pure neutron matter ($\delta=1$) appears to be 
unbound. 
The difference between the E/A values corresponding to the symmetry parameters 
$\delta=0$ and $\delta=0.2$ remains almost constant for $\r_v >0.4$ fm$^{-3}$.
This fact is reflected in a strong change in the trend of the 
symmetry energy parameter $a_4$  as a function of the density for $\r_v >0.4$ fm$^{-3}$.
This can be seen clearly in Fig. 5, where we have represented 
the $a_4$ parameter in the ZRL1 set as a function of the density. 
The strong deviation of $a_4$ from 
the linear dependence at very high densities is related to the highly 
increasing role of the NLSI terms, especially the quartic ones, in this region. 
Anyway, this set generates reasonable solutions
up to densities larger than the HFSI set does. In this latter case,
to get a value of $K=250$ MeV, one needs a contribution of the
NLSI terms producing a rapid decreasing of $m^*_\s$
with the density for $\rho_v>\rho_0$ [14]. 
Thus, $m^*_\s$ becomes almost zero for densities larger 
than  $\rho_v\simeq 0.22$ fm$^{-3}$. That is why the quantities 
E/A and $a_4$ have only been represented up to this density in 
Figs. 1 and 5.

For comparison, we have also represented in Fig. 5 with the NLHF label the results 
for $a_4$ obtained in Ref. [10], within an approximate procedure to 
include exchange contributions of the NLSI terms of scalar type.
One can see from this figure that the results corresponding to the NLHF approximation 
lie significantly below our results. However, the NLHF approximation does not include 
isovector mesons. Of course, the inclusion of these mesons would significantly increase 
the value of $a_4$.

We have mentioned above that the NLSI terms are explicitly isovector-independent [see Eq. (1)]. 
However, their contributions to the nucleon self-energies manifest
a strong isovector structure due to the exchange part.

To illustrate this point, we show in Figs. 6-9 the density dependence of 
the quantities $\Si^{(3)}_{sE}$, $\Si^{(4)}_{sE}$, $\Si^{(3)}_{0E}$, 
$\Si^{(4)}_{0E}$, respectively. 
The first two quantities correspond to the exchange (Fock) contributions of the cubic 
and quartic terms, respectively, of the NLSI part of the Lagrangian to the 
scalar nucleon self-energy, whereas the last two quantities represent the 
exchange contributions to the (time component of the) vector nucleon self-energy.
We have considered three values of the asymmetry parameter: $\delta=$0, 0.5, 1. 
Notice that for the case $\delta=1$, the proton self-energies 
correspond to a proton moving in pure neutron matter and that, in this case,
$\Si^{(3,4)}_{0E}$ are equal to zero.
 
From Figs. 6-9, it is seen that the exchange (Fock) 
self-energies $\Si^{(3,4)}_{iE}$ ($i=s,0$) represent a very important 
contribution to the respective total 
self-energies even at normal densities of NM (especially that of $\Si^{(3,4)}_{sE}$.
This contribution is appreciably smaller for the ZRL2 set with a larger value of the
compressibility modulus, due to the smaller values of the parameters $\bb$ and $\cb$.
It is also seen that $\Si^{(3,4)}_{iE}$
($i=s,0$) are essentially dependent on the neutron excess in the system
(due to the Hartree-Fock framework used).

\bigskip
\centerline{\it B. Finite nuclei}

In order to make a more complete analysis of the properties of
our nonlinear RHF model, we have carried out calculations for finite 
nuclei with the ZRL1 and the ZRL2 sets of Table I. Actually, as we have explained 
above, we have taken into account, besides the saturation NM properties, 
the experimental values of the binding 
energies, spin-orbit splittings and $r.m.s.$ charge radii of finite nuclei 
in choosing the parameters of the ZRL1 and ZRL2 sets.
We remind that the ZRL1 and ZRL2 sets contain five fitting parameters, as explained 
in subsection $A$ dedicated to NM.

In Tables II and III, we present the RHF results of our ZRL1 set\footnote{The
results for finite nuclei with the ZRL2 set are almost identical to those 
obtained with the ZRL1 set. This is due to the fact that the properties of NM
obtained with both parametrizations are very similar in the range of densities 
relevant in finite nuclei. Only for larger densities, the properties of these 
two sets are significantly different. Notice that the smaller values of 
the $\bb$ and $\cb$ coefficients in the ZRL2 set than in the ZRL1 one is related to the 
larger value of the compressibility modulus in the former case than in the second one.}, 
containing the exact exchange contribution of the NLSI, given by $U_{SI}$ in Eq. (1), 
for the ground state properties for the five doubly-magic nuclei: 
$^{16}O$, $^{40}Ca$, $^{48}Ca$, $^{90}Zr$, $^{208}Pb$.
For comparison, we also present the results of our HFSI set of Ref. [14] 
(where the exchange terms of NLSI are treated in an approximate way) and experimental values.
The C.M. values in Table II indicate an estimation for the
center-of-mass correction to the total energy,
which is not included in the ZRL1 and HFSI results.

From Tables II and III, one can see that
the ZRL1 set allows a quite good description of binding energies, spin-orbit 
splittings and $r.m.s.$ charge radii for spherical nuclei.
The results are comparable to those of the HFSI set, 
although, as we have already indicated above, the ZRL1 set has 
reasonable solutions for NM up to much higher densities than the HFSI one has. 
As is common in the RHF approaches containing the exchange of pions [9], 
the ZRL1 set also predicts a strong reduction of the spin-orbit splitting
as going from the $^{40}Ca$ nucleus to the $^{48}Ca$ one. We explained in Ref. [9]
that this fact is a consequence of the small value of the pion mass.

Figs. 10-14 show the calculated charge distributions for 
the indicated nuclei and the corresponding experimental ones for comparison.
It is remarkable that there is a very good agreement between the theoretical 
results and the experimental data, specially for the $^{16}O$ and 
$^{40}Ca$ nuclei.

%\section{CONCLUSIONS}
\section{Conclusions}

In this work, we have considered a Lagrangian including, 
beside the exchange of $\s$, $\w$, $\pi$ and $\rho$ mesons between nucleons,
NLSI of zero range, associated with three- and four-body forces, 
in the framework of the RHF approximation.
The Fierz transformation allows us to express the exchange (Fock) terms of the NLSI 
explicitly via the direct (Hartree) terms in an exact way.
The model has been applied to NM and finite nuclei.

The NLSI $(\pb\p)^3$ and $(\pb\p)^4$ considered here are constructed from the scalar 
densities. However, the relativistic Hartree-Fock procedure generates, in 
this case, the nucleon self-energies (in the respective single-particle 
Dirac equation) with the space-time transformation properties of a 
relativistic scalar, vector and tensor (the pseudoscalar and axial 
components do not survive).
It should be mentioned also that the NLSI generate strong density dependence 
of the respective nucleon self-energies $\Si_{sq}^{(3,4)}$, $\Si_{vq}^{(3,4)}$, and $\Si_{Tq}^{(3,4)}$. 
In the RHF framework, the method suggested manifests also a 
strongly developed ISVS of the self-energies generated by 
isovector-independent NLSI, this point being one of the basic 
features of the RHF framework 
(notice that in the relativistic Hartree approximation there is no ISVS due to
the NLSI, neither, in the usual form $\st$- and $\s^4$ nor
in the ZRL).

Let us mention that the same method, as suggested in the present paper, can 
be utilized to treat the self-interactions of the $\w^4$-type ($\w$ being
the vector-isoscalar meson field) and also associated with four-body 
forces [15,17]. In this case, in the ZRL, the nonlinearities are 
constructed by mean of the relativistic vector 
densities $\pb\gamma^\mu\p$ and contain eight fermion operators. 
Self-interactions of the $\w^4$-type will introduce an additional fitting parameter. 
We shall consider this case in more detail in a forthcoming paper.
Actually, NLSI of different types ($\s\w^2$, $\s^2\w^2$, etc. [15]) can be 
treated in the ZRL in the same fashion as we did for the $U_{SI}$ term.

The results given in sect. III show that our model with five 
parameters is flexible enough to generate NM properties, in the range of densities relevant
for finite nuclei, compatible with data inferred from nuclear experiments. 
Although the present model allows to get reasonable results 
for NM until values of the density much higher than 
those of the HFSI set of Ref. [14], we cannot expect that our results, obtained 
with a parametrization that describes adequately finite nuclei properties, 
could be considered realistic for high densities, for example, larger than 
2$\r_0$.

The calculations for doubly-magic nuclei have been also carried out 
within the ZRL1 set. Our results show a rather good agreement between 
theory and experiment for binding energies, spin-orbit splittings, $r.m.s.$ 
charge radii and charge densities, especially, in light and mid-weight nuclei.

\bigskip
\no
%\c{\bf ACKNOWLEDGEMENTS}
\c{\bf Acknowledgements}

Two of the authors (L.N.S. and V.N.F.) are grateful to the University of 
Cantabria for hospitality.
This work has been supported by the Ministerio de Ciencia y Tecnolog\'\i a
 grand BFM2001-1243.

%\newpage

\section{References}
\baselineskip=0.3 cm
\begin{enumerate}
\item R. Brockmann, Phys. Rev. C{\bf 18}, 1510 (1978).
\item C. J. Horowitz and B. D. Serot, Nucl. Phys. {\bf A399}, 529 (1983).
\item A. Bouyssy, J.-F. Mathiot, N. V. Giai and S. Marcos, 
Phys. Rev. C{\bf 36}, 380 (1987).
\item P. G. Blunden and M. J. Iqbal, Phys. Lett. 196B, 295 (1987).
\item J.-K. Zhang and D.S. Onley, Phys. Rev. C{\bf 44}, 1915 (1991).
\item R. Fritz and  H. M\"uther, Phys. Rev. C{\bf 49}, 633 (1994).
\item H. F. Boersma and R. Malfliet, Phys. Rev. C {\bf 49}, 233 (1994).
\item Zhong-yu Ma, Hua-lin Shi and Bao-qiu Chen, Phys. Rev. C {\bf 50}, 144 (1995).
\item M. L\'opez-Quelle, N. Van Giai, S. Marcos and L. Savushkin,
Phys. Rev. C {\bf 61}, 064321 (2000)
\item V. Greco, F. Matera, M. Colonna, M. Di Toro, G. Fabbri, Phys. Rev. C{\bf 63}, 035202 (2001).
\item J. Boguta and A. R. Bodmer, Nucl. Phys. {\bf A292}, 413 (1977). 
\item A. Bouyssy, S. Marcos and Phan Van Thieu, Nucl. Phys. {\bf A422}, 541 (1984).
\item P. Ring, Prog. Part. Nucl. Phys. {\bf 37}, 193 (1996).
\item P. Bernardos, V. N. Fomenko, N. V. Giai, M. L\'opez-Quelle,
S. Marcos, R. Niembro and L. N. Savushkin, Phys. Rev. C {\bf 48}, 2665 (1993).
\item L. N. Savushkin, S. Marcos, M. L\'opez-Quelle, P. Bernardos,  
V. N. Fomenko and R. Niembro, Phys. Rev. C {\bf 55}, 167 (1997).
\item J. A. Maruhn, T. B\"urvenich and D. G. Madland, J. Comput. Phys. {\bf 
169}, 238 (2001).
\item T. Hoch, D. Madland, P. Manakos, T. Mannel, B. A. Nikolaus and
D. Strottman, Phys. Rep. {\bf 242}, 253 (1994).
\item T. B\"urvenich, D. G. Madland, J. A. Maruhn and P.-G. Reinhard,
arXiv: nucl-th/0111012.
\item A. Sulaksono, T. B\"urvenich, J. A. Maruhn and P.-G. Reinhard and W. Greiner
arXiv: nucl-th/0301072.
\item G. A. Lalazissis, J. K\"onig and P. Ring, Phys. Rev. C {\bf55}, 540, (1997).
\item B. Frois, Proceedings of the International Conference on Nuclear Physics, 
Florence, Italy, 1983.

\end{enumerate}

%\newpage

\vspace{4cm}

%\newpage

\oddsidemargin-0.2cm
%\centerline{\bf TABLES}
\bigskip
%\section*{TABLE I}
%{\bf TATBLE I}
\small
%\vspace {2cm}
\begin{table}[ht]
%\section*{TABLE I}
{\bf Table 1}. Adjusted parameters and some properties of symmetric NM 
for the parameter sets ZRL1, ZRL2, 
corresponding to the zero range limit (ZRL) considered here, and HFSI from Ref. [14]. 
The value of the effective nucleon mass $M^*$ is defined in the 
conventional way and is given at the Fermi surface. $a_4$ is the 
symmetry energy parameter. The equilibrium density $\rho_0$ is given in $fm^{-3}$, 
the mass of the scalar meson $m_\sigma$, the energy per particle at the equilibrium $E/A$, 
the compressibility modulus $K$ and the symmetry energy parameter $a_4$ are given in MeV. 
\medskip

\centering
\begin{tabular}{@{\extracolsep{0.5mm}}lccccccccccc}
%\hline
   Set & \mc{c} {$\gs2/4\pi$} & \mc{c}{$\gw2/4\pi$} & 
\mc{c}{$m_\s $} & \mc{c}{$\bb*10^3$} & \mc{c}{$\cb*10^3$} & 
\mc{c}{$\rho_0 $} & \mc{c}{$E/A$} & \mc{c}{$K$} & 
\mc{c}{$M^*/M$} &\mc{c}{$a_4$}\\
\hline
ZRL1  & 5.5743 &11.732&497.8& 2.9646 & 51.00 &0.155& -16.39 & 250 & 0.58 & 35.0 \\
ZRL2  & 5.5121 &11.565&496.8& 2.4591 & 46.50 &0.154& -16.37 & 275 & 0.58 & 35.0 \\
HFSI  & 4.005  &10.4  &412.0& -6.718 &-14.61 &0.140& -15.75 & 250 & 0.61 & 35.0 \\
%\hline 
\end{tabular}
\end{table}

%\pagebreak
\vspace{2cm}

\oddsidemargin-0.2cm
\small
%\vspace {2cm}
\begin{table}[ht]
%\section*{TABLE II}
{\bf Table 2}. Comparison of the results of the present 
calculation for the ZRL1 set in finite nuclei with
the corresponding results of the HFSI set of Ref. [14] and the experimental ones. 
The total binding energy per particle E/A, 
the non-relativistic center-of-mass correction to E/A, 
and the proton spin-orbit splitting $\Delta_{LS}$ for
the shells $1p$ of the $\O$ nucleus and $2d$ of the $^{40}Ca$ and $^{48}Ca$ nuclei
are given in MeV (the experimental values of $\Delta_{LS}$ 
in the $Ca$ isotopes are not very well established [9]). 
The $r.m.s.$ charge radii $r_c$ are given in fm. 

\medskip
\centering
\begin{tabular}{@{\extracolsep{-2.mm}}lccccccccccccccccc}
%\hline
   & \mct{c}{\sa}  & & \mct{c}{\sb}  & & \mct{c}{\sc}   & & 
\mcd{c}{\sd}  & & \mcd{c}{\se}\\
     \cline{2-4}       \cline{6-8}       \cline{10-12}
\cline{14-15}     \cline{17-18}
Set & \mc{c}{$-E/A$}&\mc{c}{$r_c$}&\mc{c}{$\Delta_{LS}$}& &
     \mc{c}{$-E/A$}&\mc{c}{$r_c$}&\mc{c}{$\Delta_{LS}$}& &
     \mc{c}{$-E/A$}&\mc{c}{$r_c$}&\mc{c}{$\Delta_{LS}$}& &
     \mc{c}{$-E/A$}&\mc{c}{$r_c$}& &
     \mc{c}{$-E/A$}&\mc{c}{$r_c$}\\
\hline
ZRL1    &7.37&2.71&6.3& &8.33&3.44&7.1& &8.51&3.49&2.6& &
         8.67&4.25& & 7.85 & 5.49 \\
HFSI    &7.43&2.73&6.4& &8.33&3.48&7.05& &8.45&3.48&3.27& &
         8.58&4.26& &7.78&5.52\\
EXP     &7.98&2.73&6.3& &8.55&3.48&6-7.6& &8.67&3.47&5& &
         8.71&4.27& &7.87&5.50\\
C.M.    &0.61&    &   & &0.20&    &   & &0.18&    &   & &
         0.08&    & &0.02&   \\
%\hline 
\end{tabular}
\end{table}

\pagebreak

\renewcommand{\baselinestretch} {1.34}
\oddsidemargin0.0cm
\small
%\vspace {2cm}
\begin{table}[ht]
%\section*{TABLE III}
{\bf Table 3}. The single particle energies (in MeV) for protons and neutrons in 
the $\O$, $^{40}Ca$, and $^{48}Ca$ nuclei. For each state, the first and the second rows
correspond to sets ZRL1 of the present paper and HFSI of Ref. [14],
respectively. References to the experimental data are given in Refs.
[9] and [14].

\medskip
\centering
\begin{tabular}{@{\extracolsep{-1.5mm}}lcccccccccccccccc}
%\hline
  & \mcq{c}{\sa}  & & \mcq{c}{\sb}  & & \mcq{c}{\sc}   \\
    \cline{2-5}       \cline{7-10}       \cline{12-15}
  & \mcd{c}{$protons$}&\mcd{c}{$neutrons$}& &
     \mcd{c}{$protons$}&\mcd{c}{$neutrons$}& &
     \mcd{c}{$protons$}&\mcd{c}{$neutrons$}      \\
     \cline{2-3}         \cline{4-5}       
     \cline{7-8}         \cline{9-10}       
     \cline{12-13}       \cline{14-15}
  State& \mc{c}{$calc.$}&\mc{c}{$exp.$}&\mc{c}{$calc.$}&\mc{c}{$exp.$}& &
     \mc{c}{$calc.$}&\mc{c}{$exp.$}&\mc{c}{$calc.$}&\mc{c}{$exp.$}& &
     \mc{c}{$calc.$}&\mc{c}{$exp.$}&\mc{c}{$calc.$}&\mc{c}{$exp.$}\\
\hline
1s$_{1/2}$ &39.51&40$\pm$8&43.89&47& &48.71&50$\pm$11&57.22& & &54.22&55$\pm$9&
59.01& \\
           &38.96&        &43.15&  & &48.64&         &56.83& & &55.3 &     &
58.8 & \\ 
           &    &        &   &   & &    &     &   &   & &   &    &   & \\
1p$_{3/2}$ &19.28&18.4 &23.36&21.8& &33.39&34$\pm$6&41.53&   & &39.33&35$\pm$7&
43.08& \\
           &18.80&     &22.73&    & &32.2 &        &40.08&   & &39.7&    &
41.95& \\
           &    &        &   &   & &    &     &   &   & &   &    &   & \\
1p$_{1/2}$ &12.98&12.1&16.95&15.7& &29.43&34$\pm$6&37.47&   & &38.24&35$\pm$7&
41.24& \\
           &12.40&    &16.22&    & &27.8 &        &35.56&   & &38.0 &     &
39.60& \\
           &    &   &    &   & &    &     &   &   & &   &    &  &  \\
1d$_{5/2}$ &    &   &    &   & &17.22&14.3-16&24.98&21.9& &23.04&20.5&26.20&16 \\ 
           &    &   &    &   & &16.2 &    &23.77&    & &23.2 &  &25.0 & \\
           &    &        &   &   & &    &     &   &   & &   &    &   &   \\
2s$_{1/2}$ &    &   &    &   & &8.64&10.9&16.36&18.2 & &15.30&15.8&18.42&12.4\\
           &    &   &    &   & &9.83&    &17.23&     & &15.8 &    &18.5 & \\
           &    &        &   &   & &    &     &   &   & &   &    &   & \\
1d$_{3/2}$ &    &   &    &   & &10.10&8.3&17.71&15.6& &20.45&15.5&21.74&12.4 \\
           &    &   &    &   & &9.19&   &16.55&    & &19.9 &    &20.2 & \\
          &    &        &   &   & &    &     &   &   & &   &    &   & \\
1f$_{7/2}$ &    &    &   &    & &    &    &    &    & &    &   &9.98&9.9 \\
           &    &    &   &    & &    &    &    &    & &    &   & 9.21&  \\
%\hline 
\end{tabular}
\end{table}

\begin{figure}
\begin{center}
\includegraphics[width=8cm,angle=270]{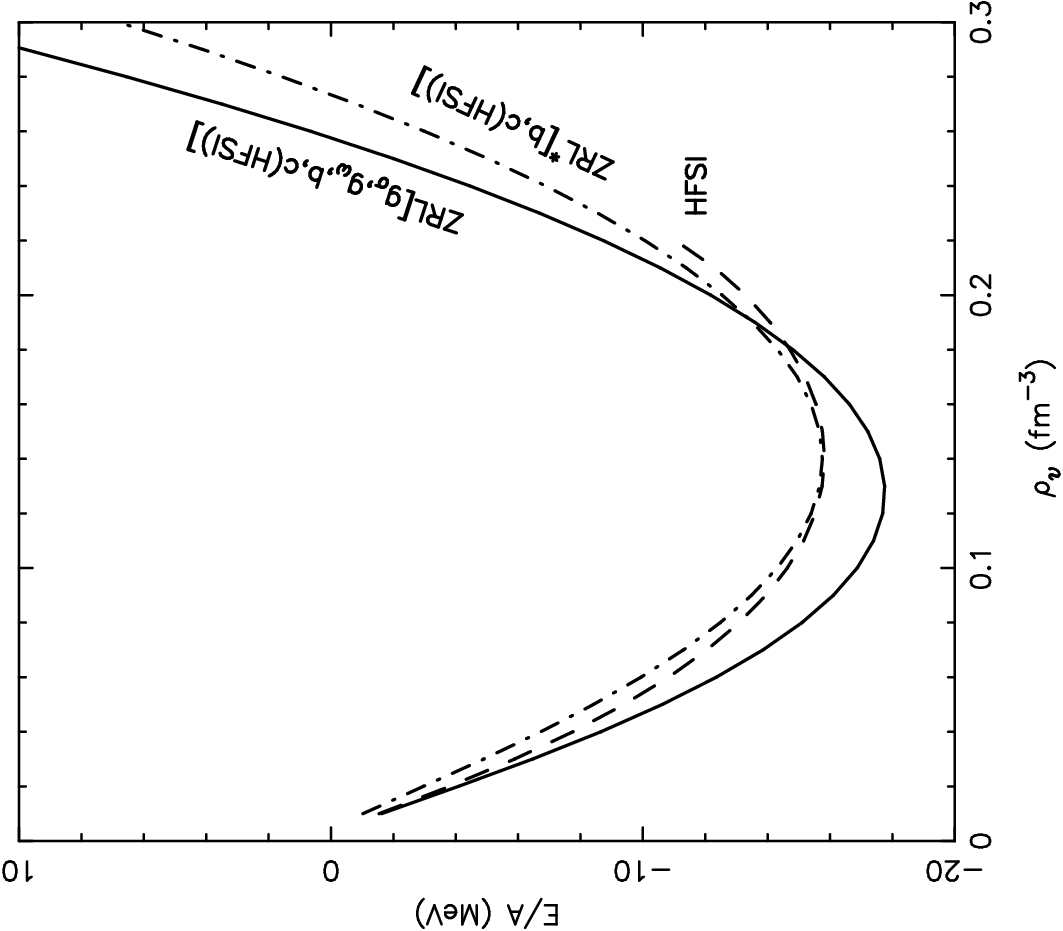}
\end{center}

%\vspace{-1 true cm}
Fig. 1: The E/A values for symmetric NM in the RHF approach in three cases:
The HFSI (dashed) curve corresponds to the results obtained with the HFSI set of Ref. [14];
the ZRL[$g_\s,g_\w$,b,c(HFSI)] (solid) curve corresponds to the 
results obtained by the method suggested in the present paper, 
with the same set of parameters as in Ref. [14]; 
and the ZRL$^*$[b,c(HFSI)] (dash-dotted) curve corresponds to the results with the 
parameters $g_\s$ and $g_\w$ chosen to get saturation at the same point
as in the HFSI set.

\end{figure}

\begin{figure}
\begin{center}
\includegraphics[width=8cm,angle=270]{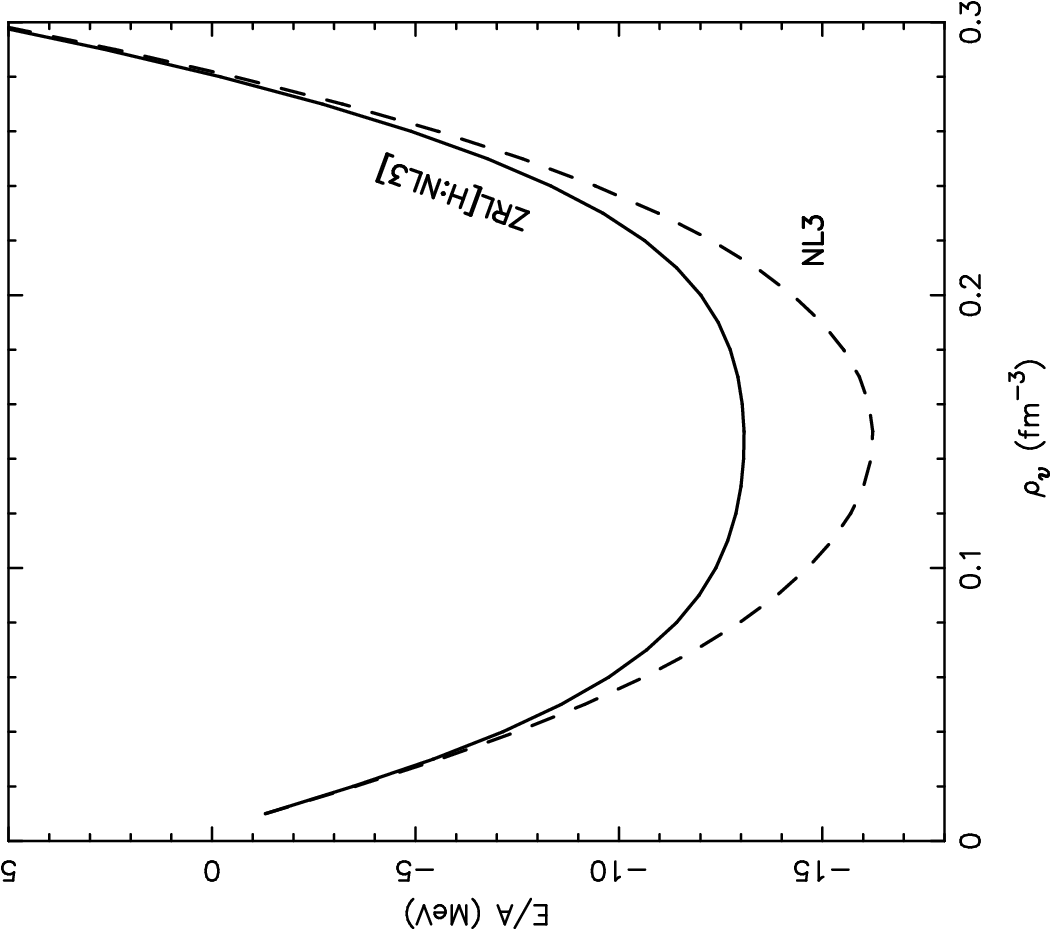}
\end{center}

%\vspace{-1 true cm}
Fig. 2. The E/A values for symmetric NM in the Hartree approach in two cases: 
The NL3 (dashed) curve represents the results obtained with the NL3 parametrization 
of Ref. [20] and the ZRL[H:NL3] (solid) curve are the results obtained with the NL3 
parametrization in the ZRL.

\end{figure}

\begin{figure}
\begin{center}
\includegraphics[width=8cm,angle=270]{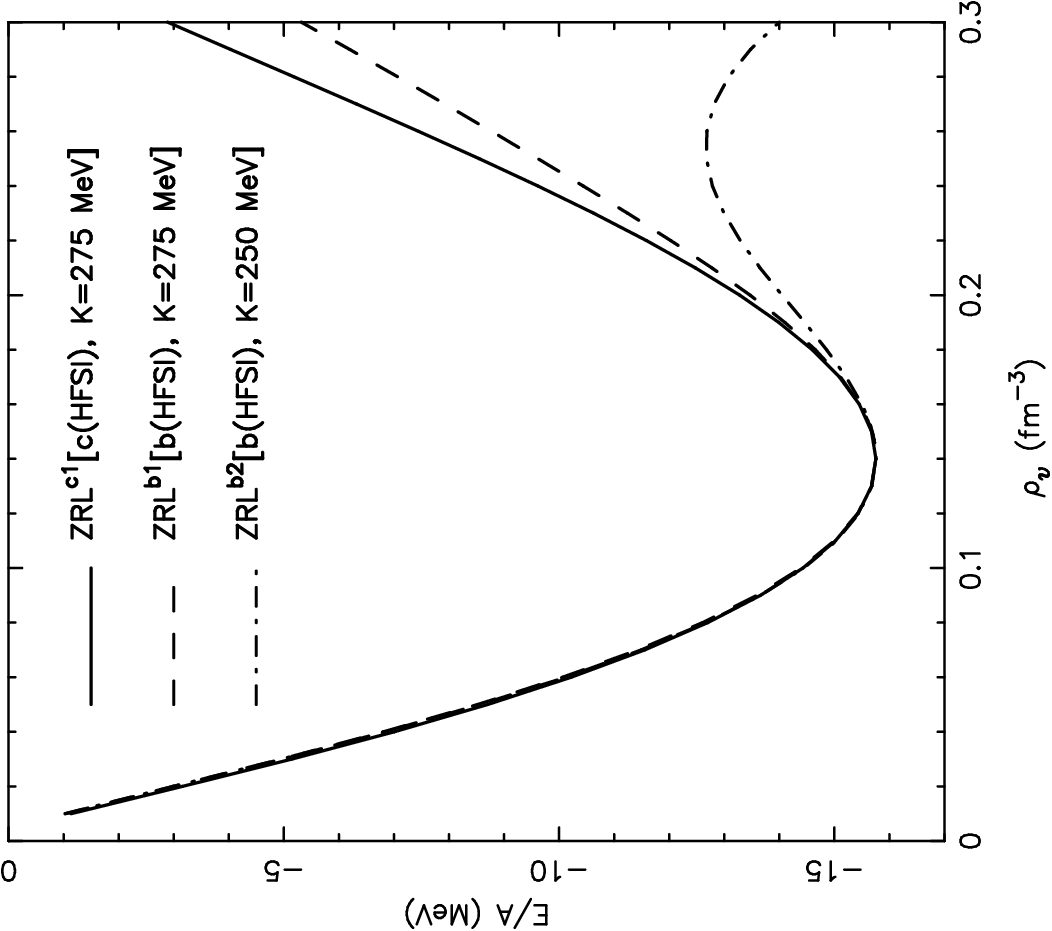}
\end{center}

%\vspace{-1 true cm}
Fig. 3. The E/A values  for symmetric NM, with the Lagrangian considered in sect. I 
(which contains the zero range NLSI given in Eq. (1))
for different choices of the  parameters
fitted to the same values of NM $\rho_0$ and $E(\rho_0)/A$ 
as in the set HFSI of Ref. [14] (see the text), 
and generating two values of the compressibility modulus: 
The ZRL$^{c1}$ (solid) curve, with $\cb<0$ taken from 
the HFSI set, corresponds to $K=275$ MeV;
the ZRL$^{b1}$ (dashed) curve, with $\bb$ taken from the HFSI set, 
corresponds to  $K=275$ MeV; and the ZRL$^{b2}$ (dash-dotted) curve, 
with $\bb$ as for the ZRL$^{b1}$ case, corresponds to  $K=250$ MeV.

\end{figure}

\begin{figure}
\begin{center}
\includegraphics[width=8cm,angle=270]{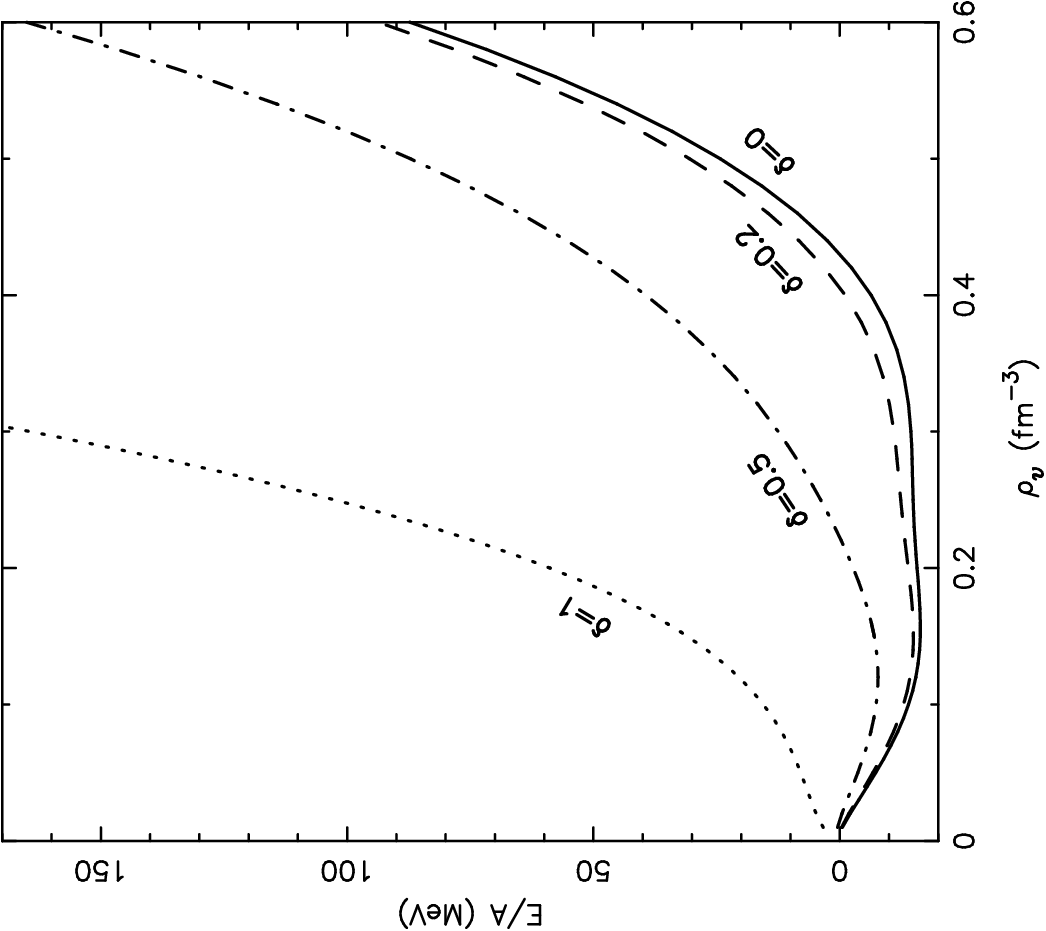}
\end{center}

%\vspace{-1 true cm}
Fig. 4. The E/A values in asymmetric NM, for the ZRL1 set, 
with different values of the asymmetry parameter $\delta$.

\end{figure}

\begin{figure}
\begin{center}
\includegraphics[width=8cm,angle=270]{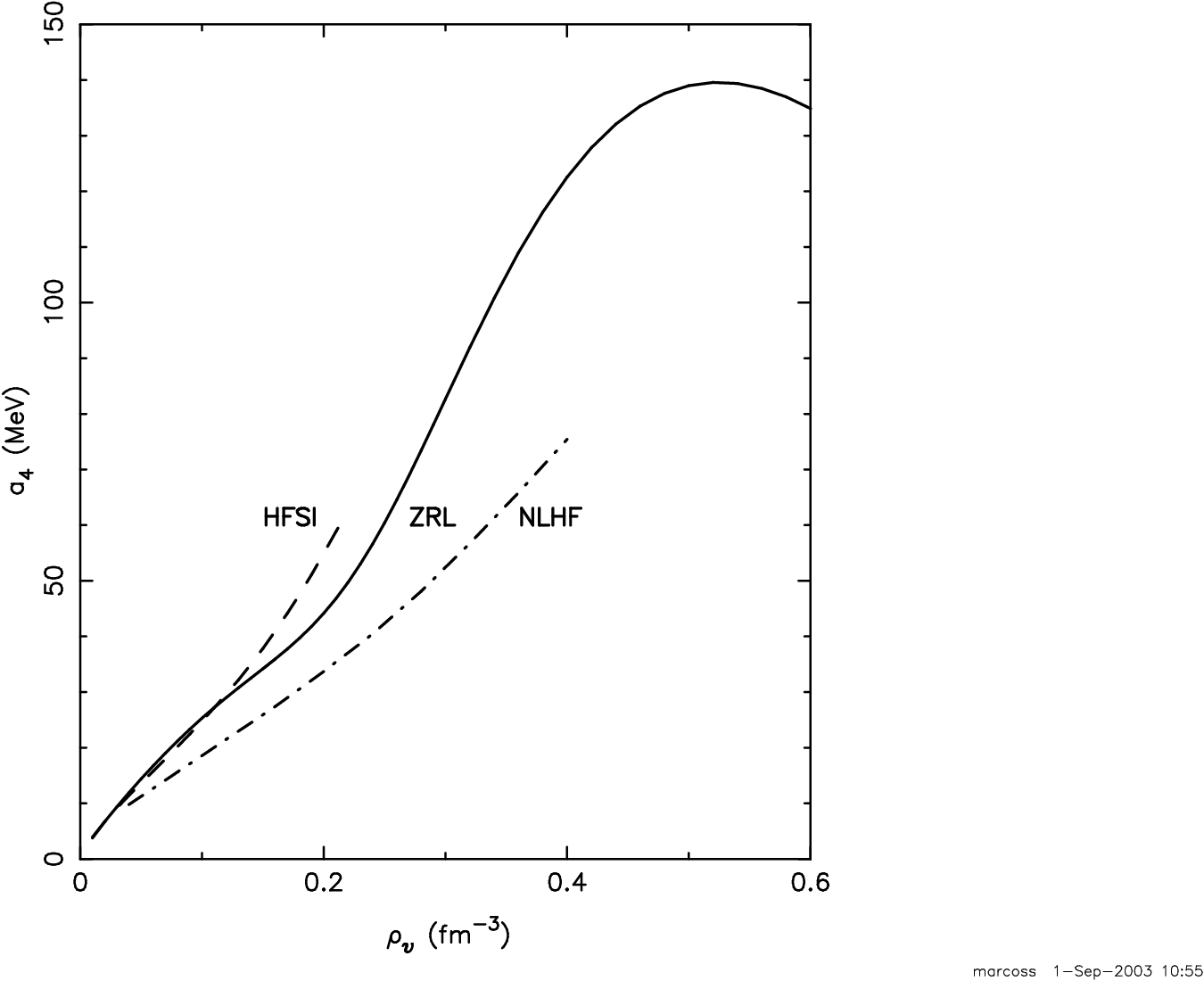}
\end{center}

%\vspace{-1 true cm}
Fig. 5. The symmetry energy parameter $a_4$ as a function of the NM
density for sets ZRL1, HFSI [14] and NLHF [10].

\end{figure}

\begin{figure}
\begin{center}
\includegraphics[width=8cm,angle=270]{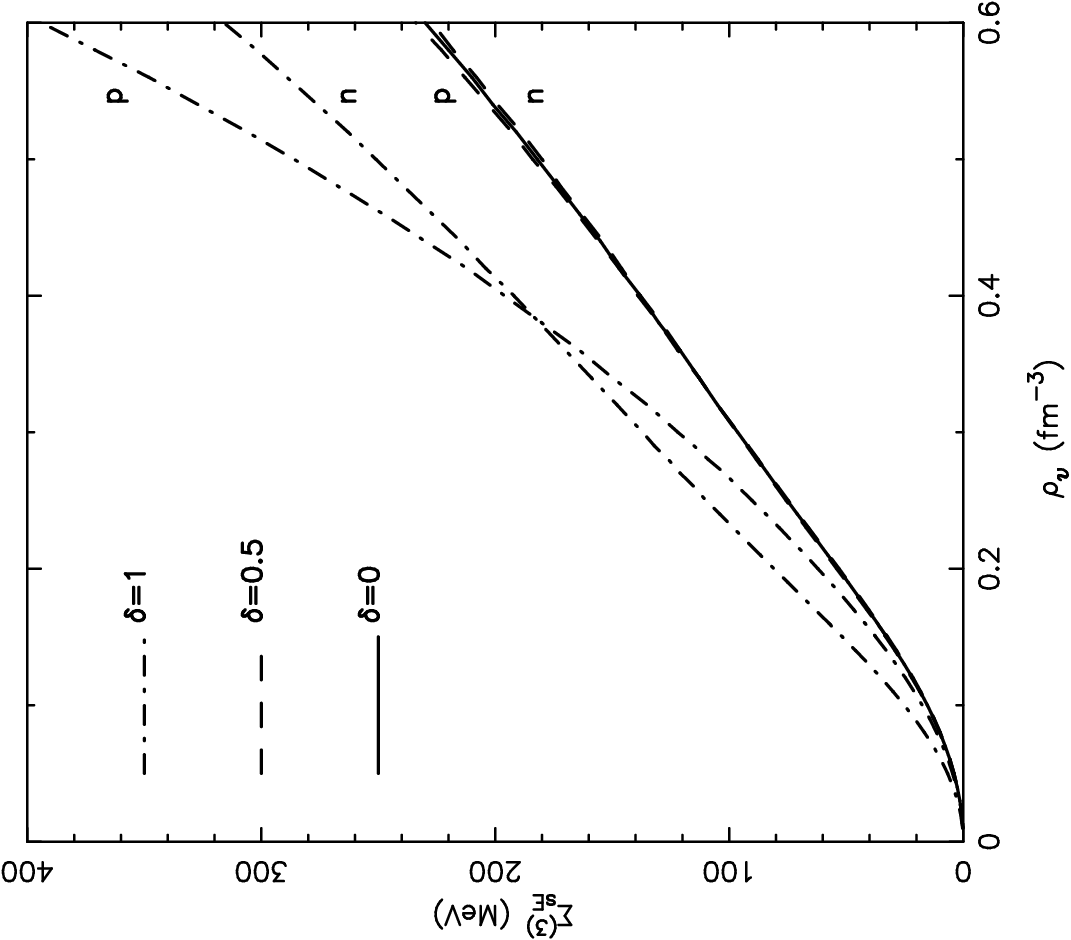}
\end{center}

%\vspace{-1 true cm}
Fig. 6. The density dependence of the exchange (Fock) 
contribution $\Si^{(3)}_{sE}$ of the $(\pb\p)^3$ term  
to the scalar nucleon self-energy for different values of 
the asymmetry parameter $\delta$. 

\end{figure}

\begin{figure}
\begin{center}
\includegraphics[width=8cm,angle=270]{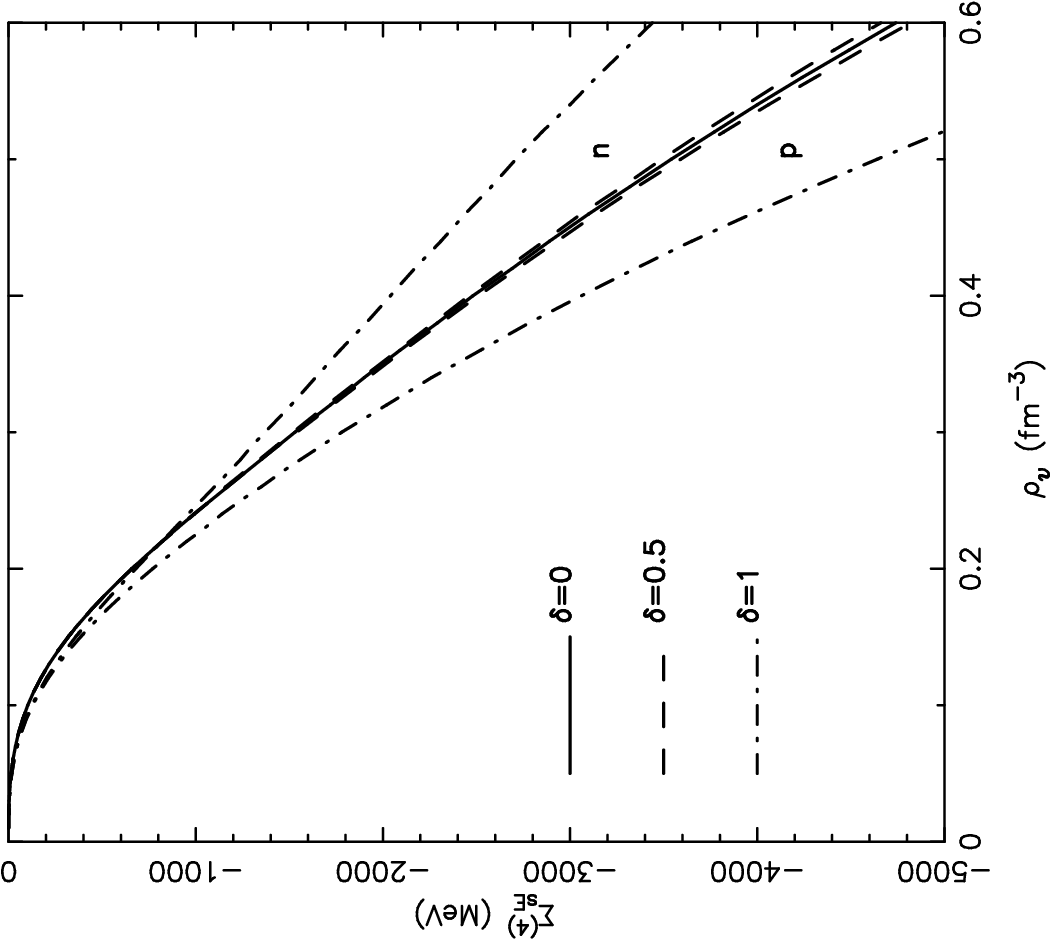}
\end{center}

%\vspace{-1 true cm}
Fig. 7. The same as in Fig. 6 but for the density dependence of the 
exchange (Fock) contribution $\Si^{(4)}_{sE}$ of the $(\pb\p)^4$ term.

\end{figure}

\begin{figure}
\begin{center}
\includegraphics[width=8cm,angle=270]{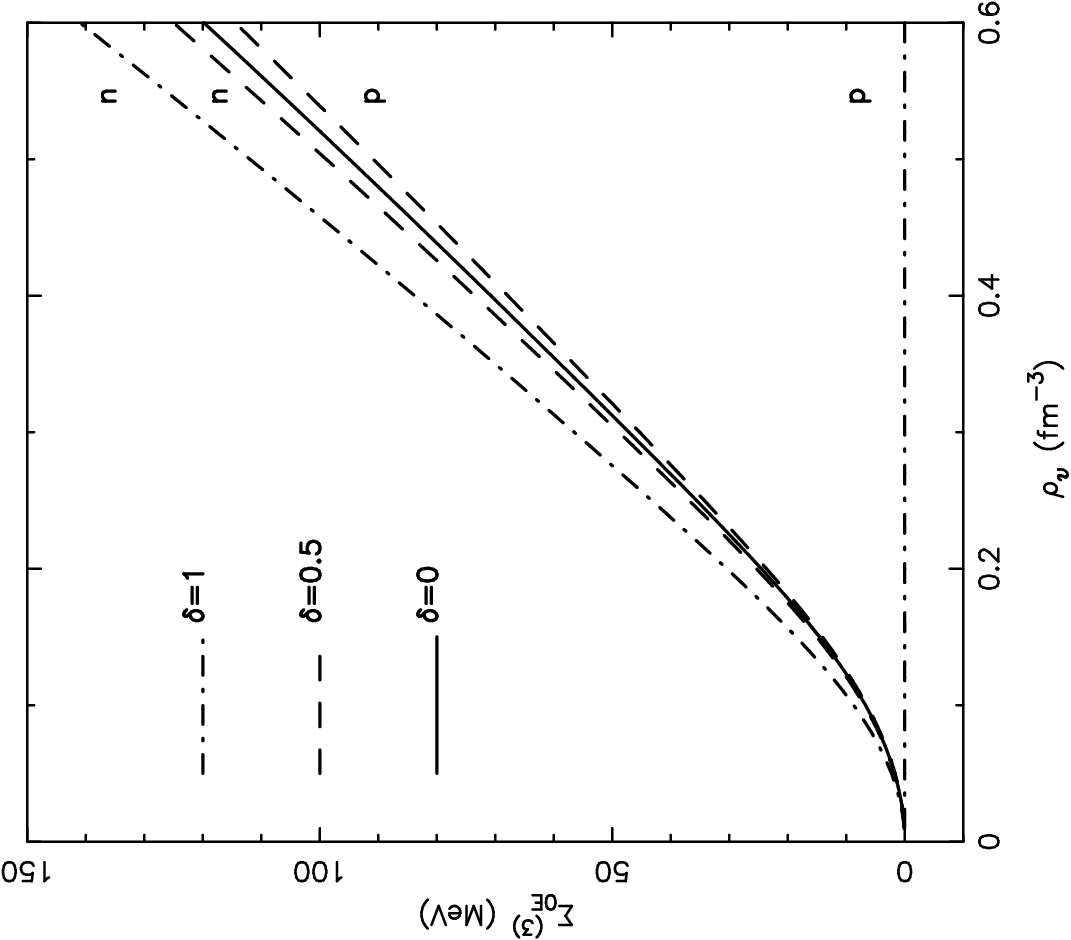}
\end{center}

%\vspace{-1 true cm}
Fig. 8. The density dependence of the exchange (Fock) 
contribution $\Si^{(3)}_{0E}$ of the $(\pb\p)^3$ term to the 
(time component of) the vector self-energy.

\end{figure}

\begin{figure}
\begin{center}
\includegraphics[width=8cm,angle=270]{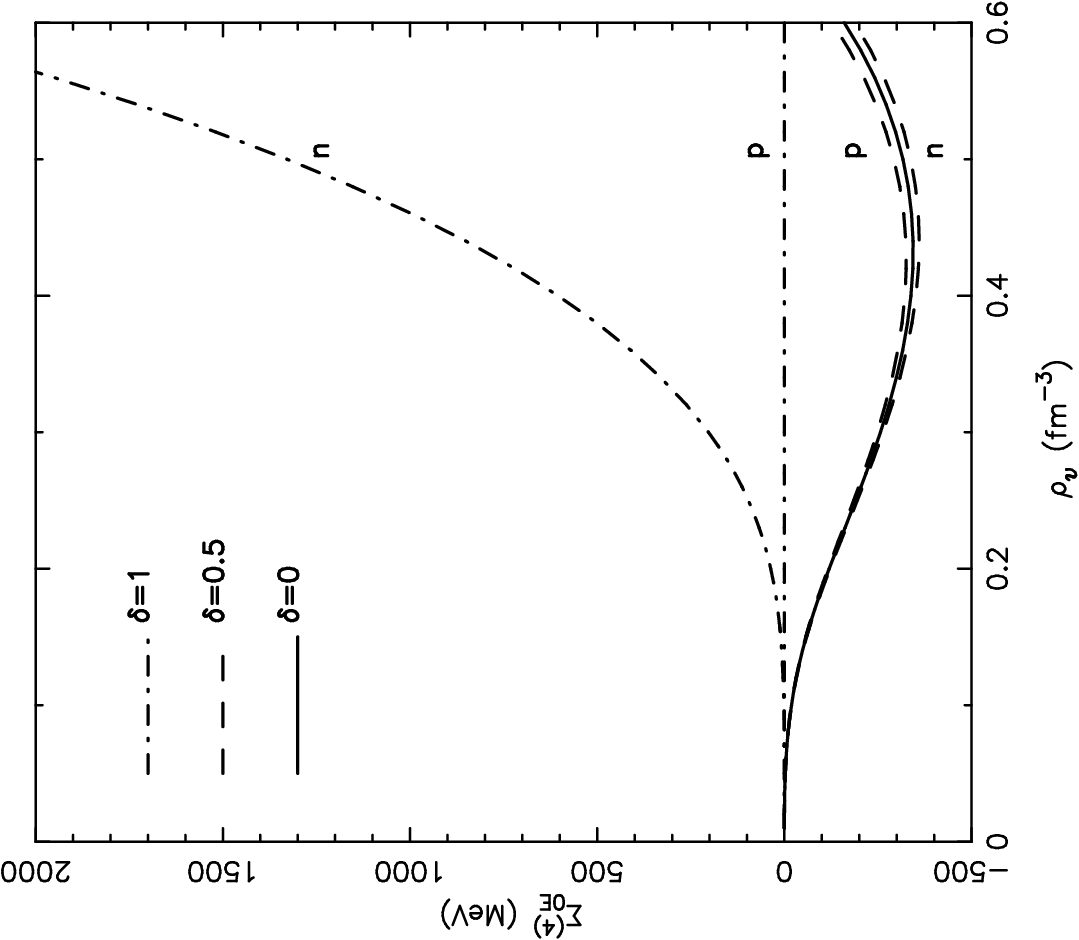}
\end{center}

%\vspace{-1 true cm}
Fig. 9. The same as in Fig. 8 but for the density dependence of the 
exchange (Fock) contribution $\Si^{(4)}_{0E}$ of the $(\pb\p)^4$ term.

\end{figure}

\begin{figure}
\begin{center}
\includegraphics[width=8cm,angle=270]{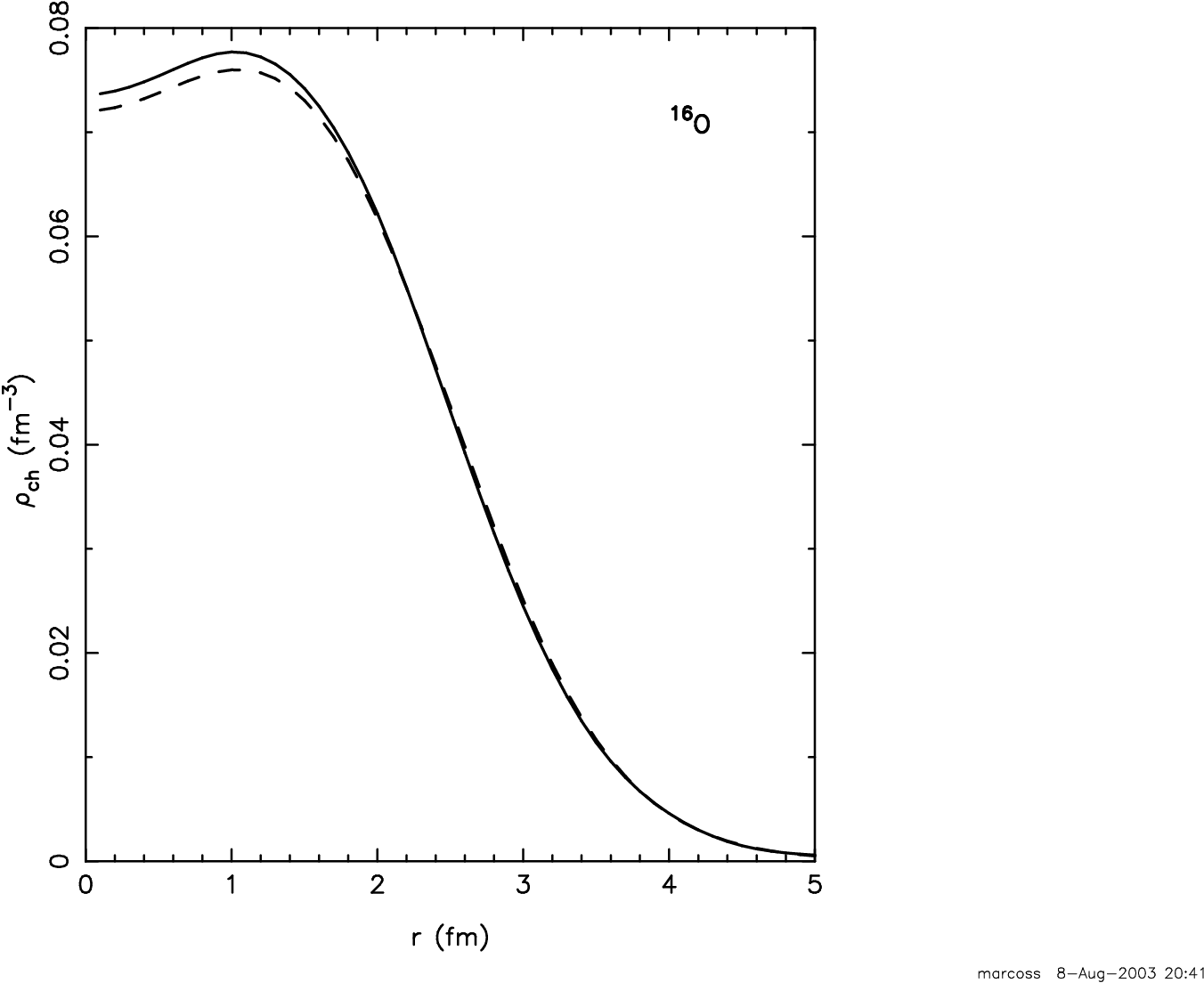}
\end{center}

%\vspace{-1 true cm}
Fig. 10. The  charge distribution (dashed line) for the $^{16}O$ nucleus, with the ZRL1 set. 
The experimental charge distribution (solid line [21]) is also shown for comparison.

\end{figure}

\begin{figure}
\begin{center}
\includegraphics[width=8cm,angle=270]{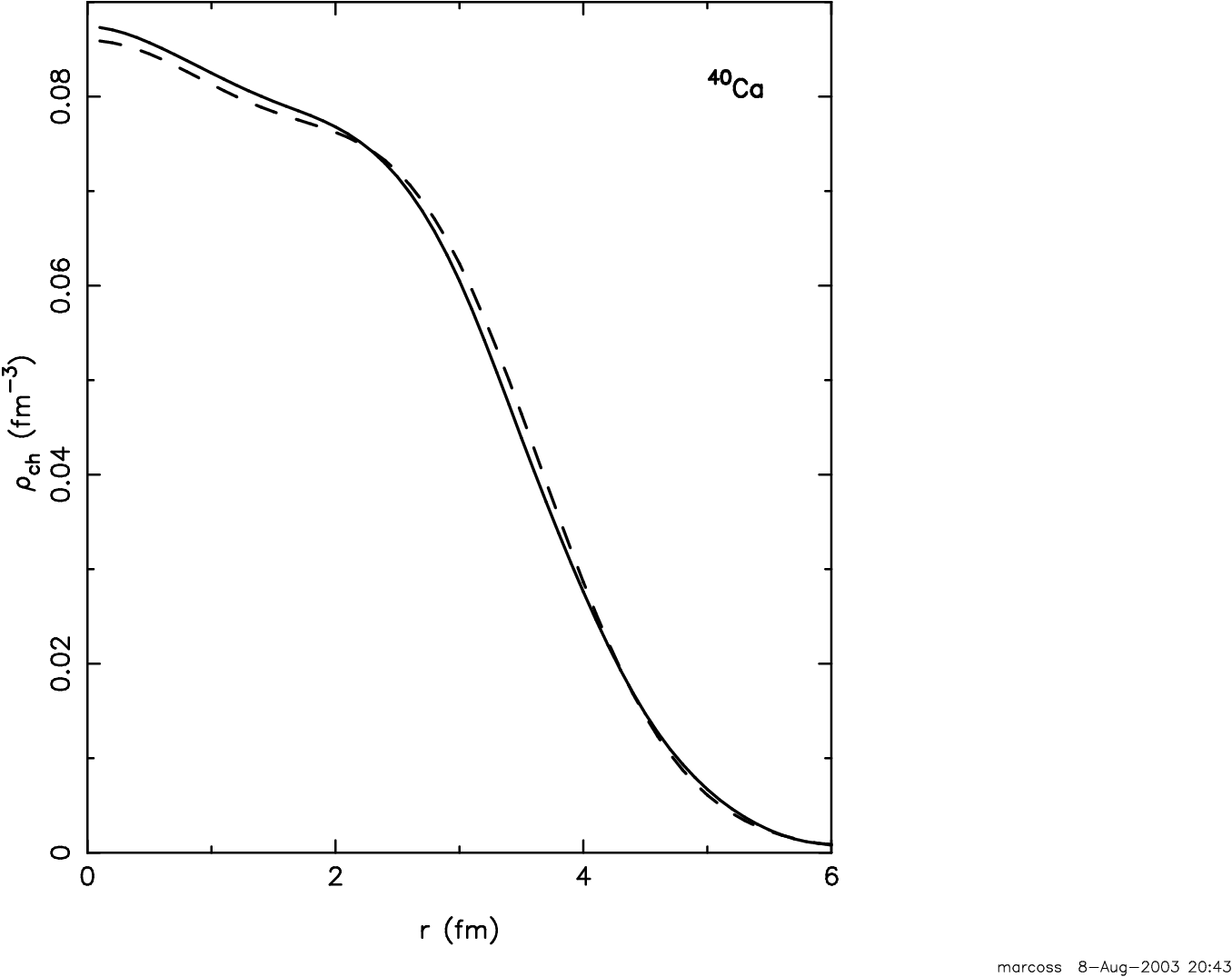}
\end{center}

%\vspace{-1 true cm}
Fig. 11. The same as in Fig. 10 but for the $^{40}Ca$ nucleus.

\end{figure}

\begin{figure}
\begin{center}
\includegraphics[width=8cm,angle=270]{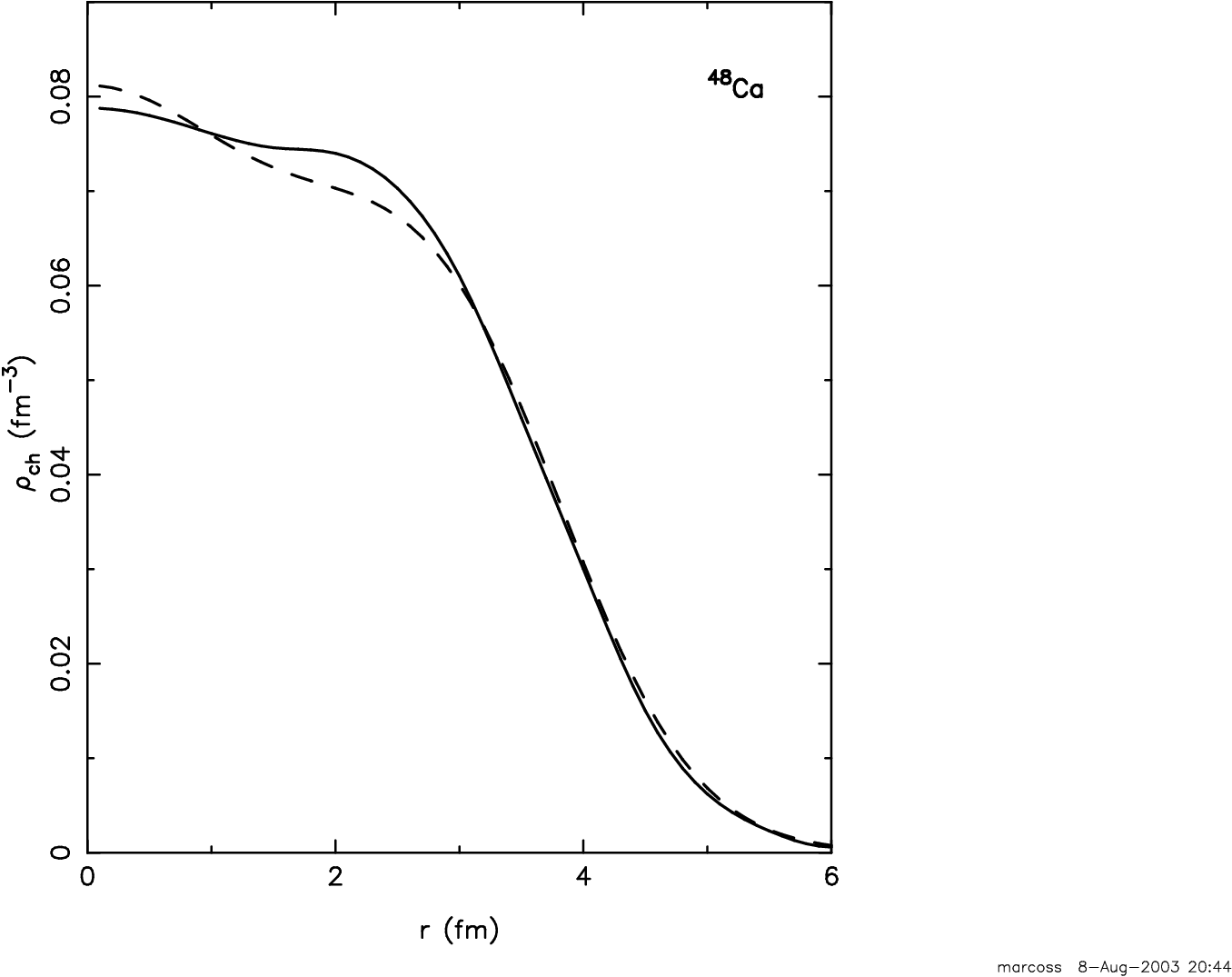}
\end{center}

%\vspace{-1 true cm}
Fig. 12. The same as in Fig. 10 but for the $^{48}Ca$ nucleus.

\end{figure}

\begin{figure}
\begin{center}
\includegraphics[width=8cm,angle=270]{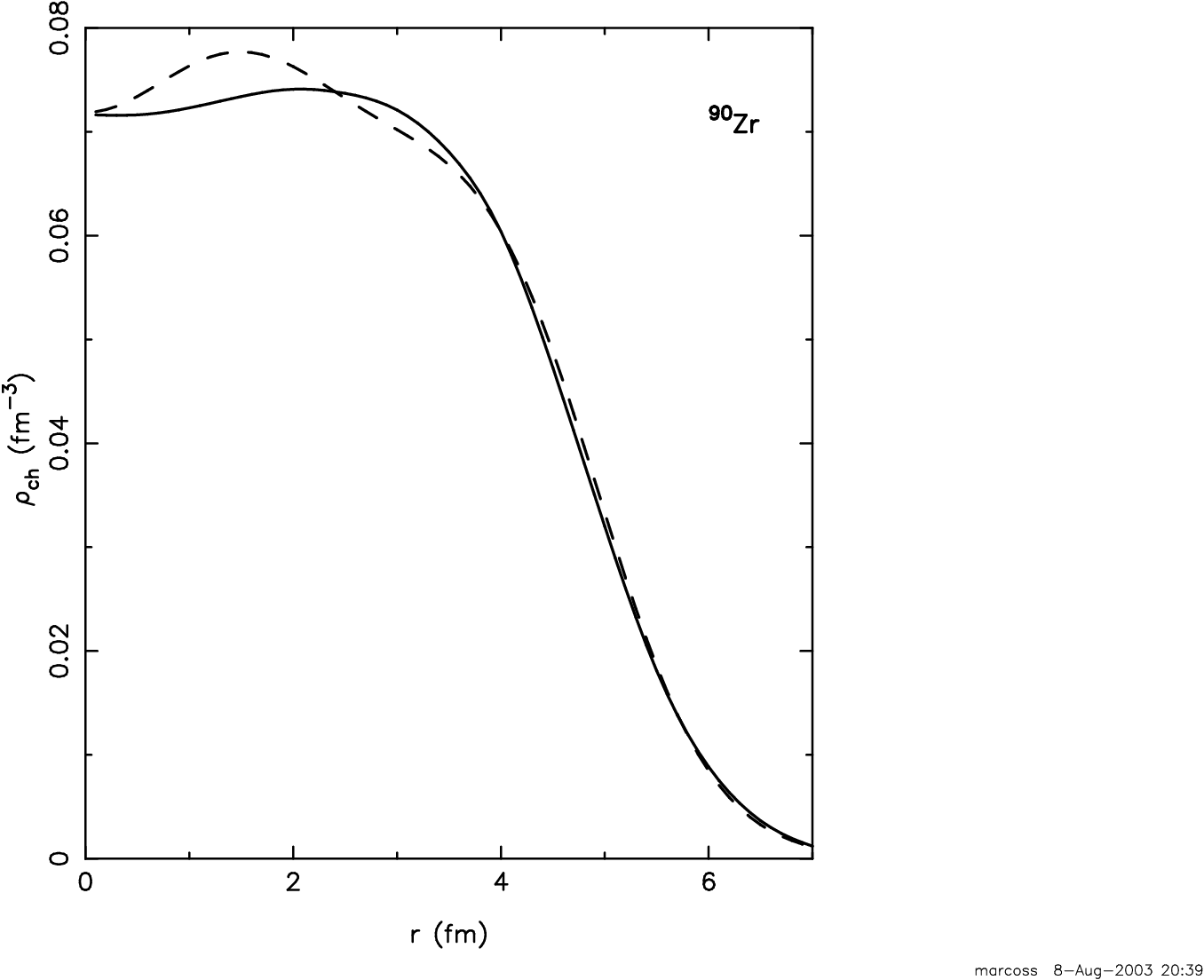}
\end{center}

%\vspace{-1 true cm}
Fig. 13. The same as in Fig. 10 but for the $^{90}Zr$ nucleus.
\end{figure}

\begin{figure}
\begin{center}
\includegraphics[width=8cm,angle=270]{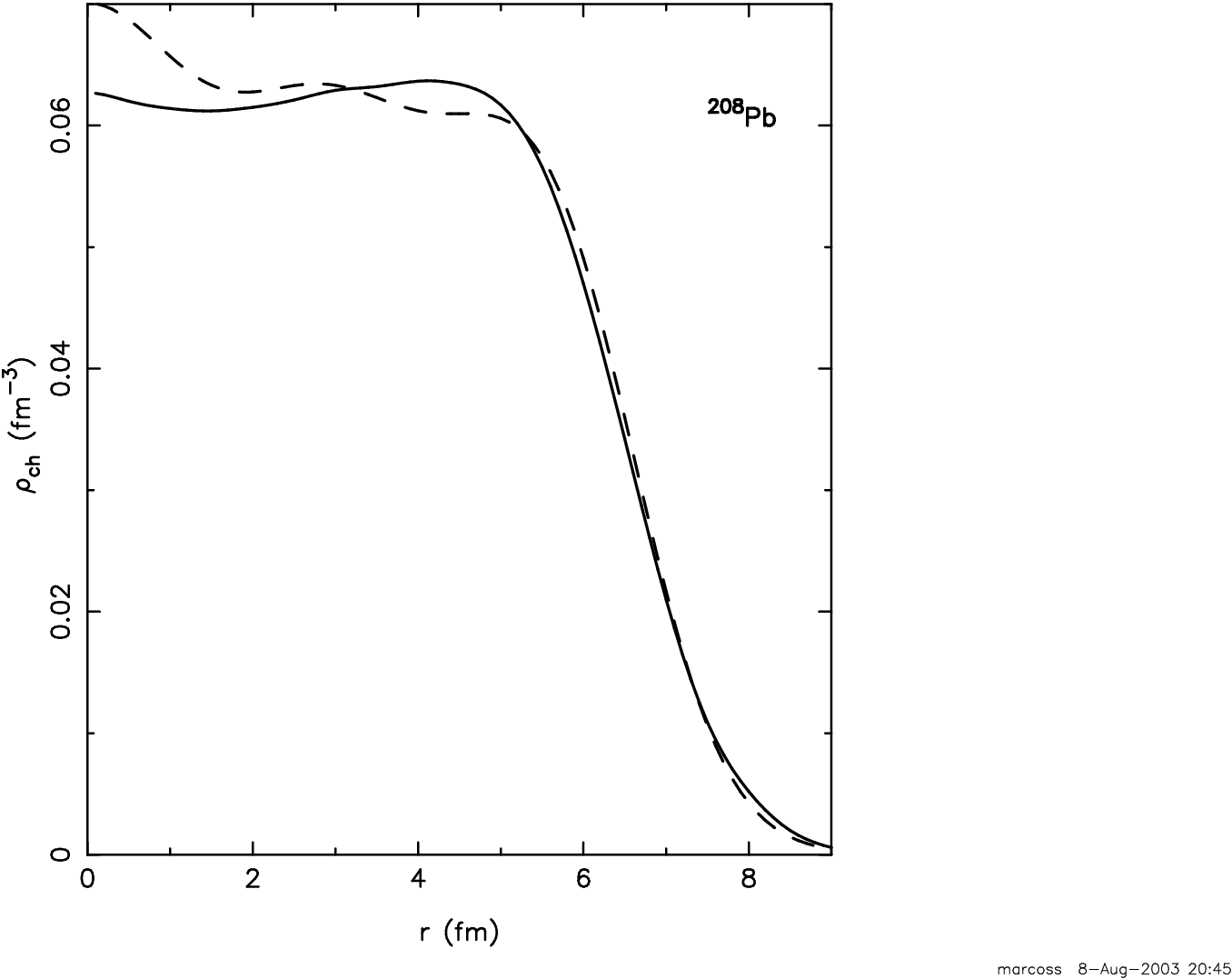}
\end{center}

%\vspace{-1 true cm}
Fig. 14. The same as in Fig. 10 but for the $^{208}Pb$ nucleus.

\end{figure}

\end{document}